\documentclass[
    ,final            % use final for the camera ready runs
    ,draft            % use draft while you are working on the paper
  ]
  {aipproc}

\layoutstyle{8x11single}

\date{}
\begin{document}

\title[\textit{The Unifying Dark Fluid Model}]{The Unifying Dark Fluid Model}


\author{Alexandre Arbey}{
  address={Universit\'e de Lyon, Lyon, F-69000, France; Universit\'e Lyon~1, Villeurbanne, \\
F-69622, France; Centre de Recherche Astrophysique de Lyon, Observatoire de Lyon, \\
9 avenue Charles Andr\'e, Saint-Genis Laval cedex, F-69561, France; CNRS, UMR 5574; \\
Ecole Normale Sup\'erieure de Lyon, Lyon, France.},
}

\begin{abstract}
The standard model of cosmology relies on the existence of two components, ``dark matter'' and ``dark energy'', which dominate the expansion of the Universe. There is no direct proof of their existence, and their nature is still unknown. Many alternative models suggest other cosmological scenarios, and in particular the {\it dark fluid} model replace the dark matter and dark energy components by a unique dark component able to mimic the behaviour of both components. The current cosmological constraints on the unifying dark fluid model is discussed, and a dark fluid model based on a complex scalar field is presented. Finally the consequences of quantum corrections on the scalar field potential are investigated.
\end{abstract}

\maketitle

%%%%%%%%%%%%%%%%%%%%%%%%%%%%%%%%%%%%%%%%%%%%
%% MAINMATTER
%%%%%%%%%%%%%%%%%%%%%%%%%%%%%%%%%%%%%%%%%%%%

\section{Introduction}
According to the standard model of cosmology, the total energy density of the Universe is presently dominated by two component densities: the {\it dark matter} component, which is a pressureless matter fluid having attracting gravitational effects, and {\it dark energy}, whose main properties are a negative pressure and a nearly constant energy density today, which evoke the idea of vacuum energy. The nature of both components remains unknown, and in the near future we can hope that the Large Hadron Collider (LHC) will be able to provide hints on the nature of dark matter. In spite of the mysteries of the dark components, it is generally considered that dark matter can be modeled as a system of collisionless particles, whereas the most usual models of dark energy are the scalar field based quintessence models. Nevertheless, many difficulties in usual dark energy and dark matter models still cast doubts on the cosmological standard model bases, leaving room for other models to be investigated. Here we consider a unifying model in which the dark matter and dark energy
components can be considered as two different aspects of a same component, the {\it dark fluid}. We will first review how such a unifying model can be constrained by the cosmological observations. Then we will consider a dark fluid model based on a complex scalar field, and we will discuss the question of the choice of the scalar field potential through an analysis of quantum corrections.

\section{Constraints on dark fluid models}
During past years, cosmological observations have greatly improved in precision and increased in number. The data analysis are generally performed within the standard model of cosmology, considering distinctly dark matter and dark energy. These results have been reinterpreted to constrain dark fluid models in \cite{Arbey:2005fn}. Furthermore, an analysis in the light of the new 5-year WMAP constraints \cite{Komatsu:2008hk} has been performed in \cite{Arbey:2008gw}. We review here some of the obtained results.

First we define the usual ratio $\Omega_D$ of the density of the dark fluid over the critical density, and $\omega_D$ the ratio of the pressure of the dark fluid over its density:
\begin{equation}
 \Omega_D = \frac{\rho_D}{\rho^c_0} \qquad,\qquad \omega_D = \frac{P_D}{\rho_D}\;\;.
\end{equation}
Recent observations of supernov\ae~of type Ia impose the following constraints on the dark fluid \cite{Arbey:2008gw}:
\begin{equation}
\Omega_D^0 = 1.005 \pm 0.006 \;\;,\;\;\omega_D^0 = -0.80 \pm 0.12 \;\;,\;\;
\omega_D^a = 0.9 \pm 0.5 \;\;,
\end{equation}
where $\omega_D$ is written in function of the expansion factor:
\begin{equation}
\omega_D= \omega_D^0 + (1-a) \omega_D^a \;\;,
\end{equation}
with $a_0=1$. The result on $\omega_D^0$ is particularly interesting, as it reveals that $\omega_D^0 > -1$, which is a property that has to be fulfilled in order for the dark fluid to be described by a scalar field.

Studying structure formation is also a useful way to constrain cosmological models. However, no stringent constraints can be determined without performing a precise analysis of a specific dark fluid model. We have nevertheless shown in \cite{Arbey:2005fn} that a dark fluid should have an equation of state respecting $\omega_D > -1/3$ at the time of structure formation.

The data on the Cosmic Microwave Background (CMB) have been greatly improved with the new WMAP 5-year data. If only the CMB data are considered, a large variety of dark fluid models would still be allowed. We can however consider that a dark fluid model is more realistic if the fluid behaves today like a dark energy with a negative pressure, but was behaving mainly like matter at the recombination time. We refer to \cite{Arbey:2008gw} for a detailed analysis of the CMB constraints.

\begin{figure}[t!]
\centering
\includegraphics[width=8.05cm]{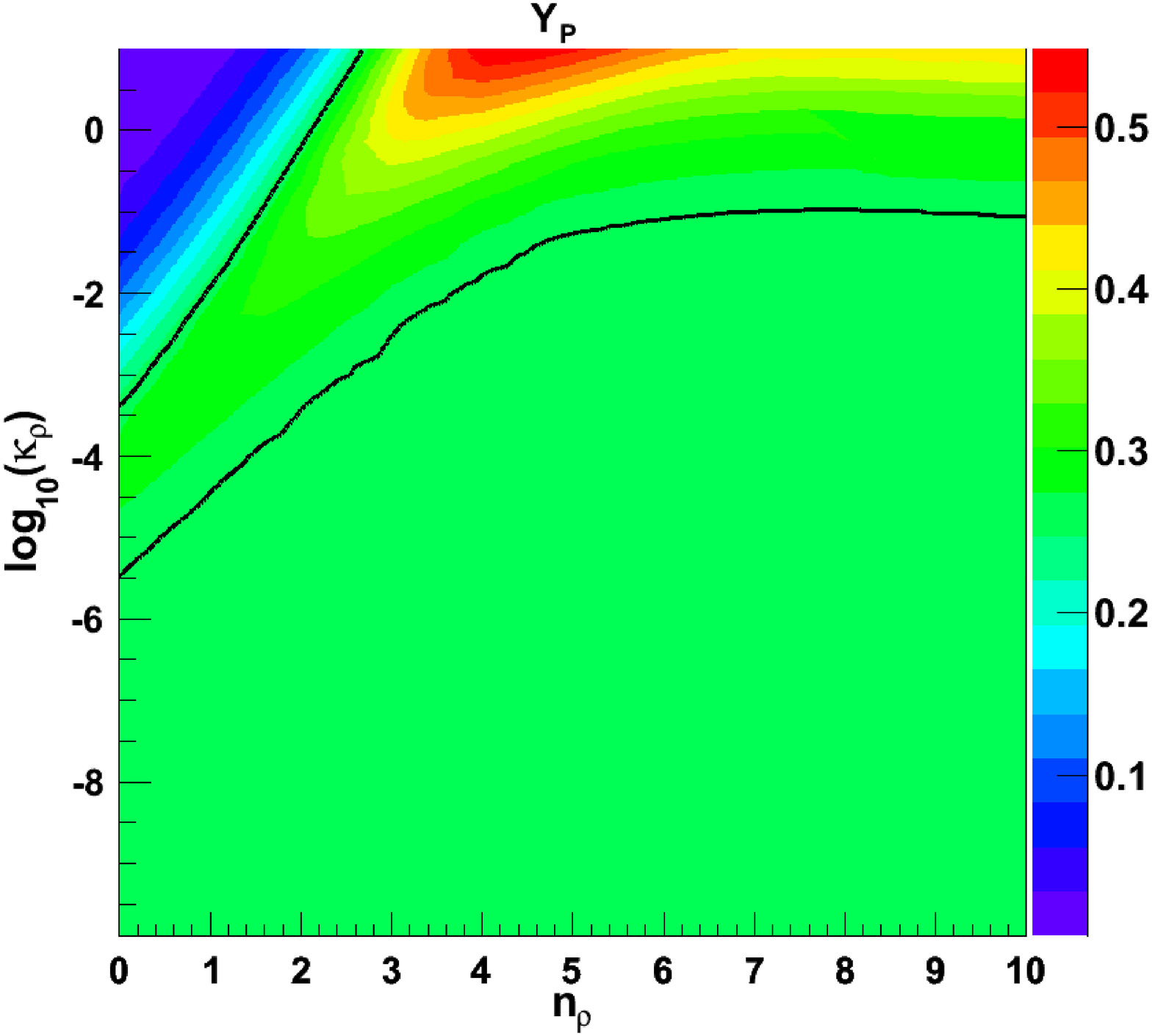}\includegraphics[width=8.05cm]{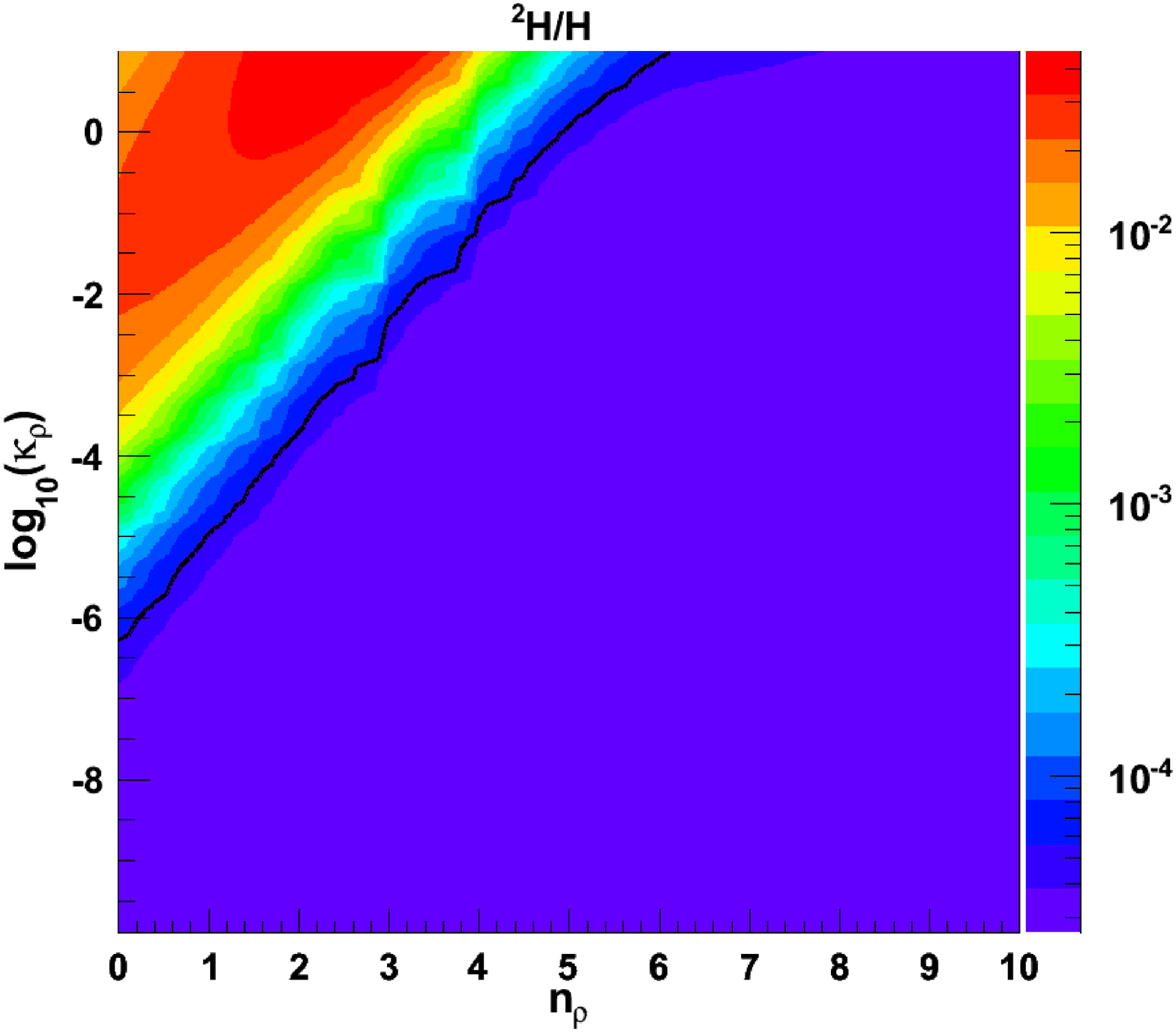}
\caption{Constraints from $Y_p$ (left) and $^2H/H$ (right) on the dark fluid parameters $(n_\rho,\kappa_\rho)$. The parameter regions excluded by BBN are located between the black lines for $Y_p$, and in the upper left corner for $^2H/H$. The colors correspond to different values of $Y_p$ and $^2H/H$.\label{BBN}}
\end{figure}

Constraints can also be derived considering primordial nucleosynthesis (BBN) models. We use the parametrization described in \cite{Arbey:2008kv} for $\rho_D$:
\begin{equation}
 \rho_D =  \kappa_\rho \rho_{rad}(T_{BBN}) \left(\frac{T}{T_{BBN}}\right)^{n_\rho} \;,\label{rhoD}
\end{equation}
where $T_{BBN}$ is the BBN temperature, $n_\rho=3(\omega_D+1)$ and $\kappa_\rho$ is the ratio of the dark fluid density over the radiation density at BBN time. In order to make a realistic analysis of the allowed cosmological scenarios, we used a modified version of the BBN abundance calculation code NUC123 \cite{Kawano:1992ua} including the parametrization (\ref{rhoD}) to compute the relevant abundances of the elements. We consider the rather conservative bounds of \cite{Jedamzik:2006xz}:
\begin{equation}
 Y_p < 0.258 \;, \qquad\qquad 1.2 \times 10^{-5} <~^2\!H/H < 5.3 \times 10^{-5} \;,
\end{equation}
for the helium abundance $Y_p$ and the primordial $^2\!H/H$ ratio. In Fig.~\ref{BBN}, the current limits on the dark fluid properties from the $Y_p$ and $^2H/H$ BBN constraints are presented. The area between the back lines on the left plot and the area on the top of the black lines on the right plot lead to unfavored element abundances.
A consequence is that if the dark fluid density were small before BBN, the dark fluid model could be compatible with BBN constraints.

Taken separately, these constraints does not seem very severe, but in a complete dark fluid scenario, all of them must be satisfied simultaneously, and when requiring that the dark fluid should also be able to explain dark matter at local scales, they become very restrictive.

In the following we will consider a dark fluid model based on a complex scalar field and confront it to the observational constraints.

\section{Complex scalar field dark fluid model}
Modeling a dark fluid is not an easy task, because such a model should describe simultaneously dark matter and dark energy properties. In the literature, only a few dark fluid type models have been investigated, the most well-known one being the Chaplygin Gas (see for example \cite{Kamenshchik:2001cp,Bean:2003ae,Bilic:2001cg}). We consider here a particular model using a scalar field, which was first introduced in \cite{Arbey:2006it}.

Dark energy shares with vacuum energy the property of having a negative pressure and a quasi-constant energy density. Therefore, a standard scalar field is often referred as a good candidate for dark energy, and models of dark energy based on scalar fields are generally referred as quintessence scalar fields \cite{Peebles:1987ek}. An important issue in these models is the choice of the potential of the scalar field, as it completely determines the behaviour of the field throughout the expansion of the Universe. Many potentials have already been investigated, and some of them, such as the decreasing exponential potential, have been already excluded by the cosmological data. However, it is still unclear how to choose the potential, but it is desirable to use potentials which can originate from physically motivated theories.

\begin{figure}[t]
\centering
\includegraphics[width=5.8cm,angle=270]{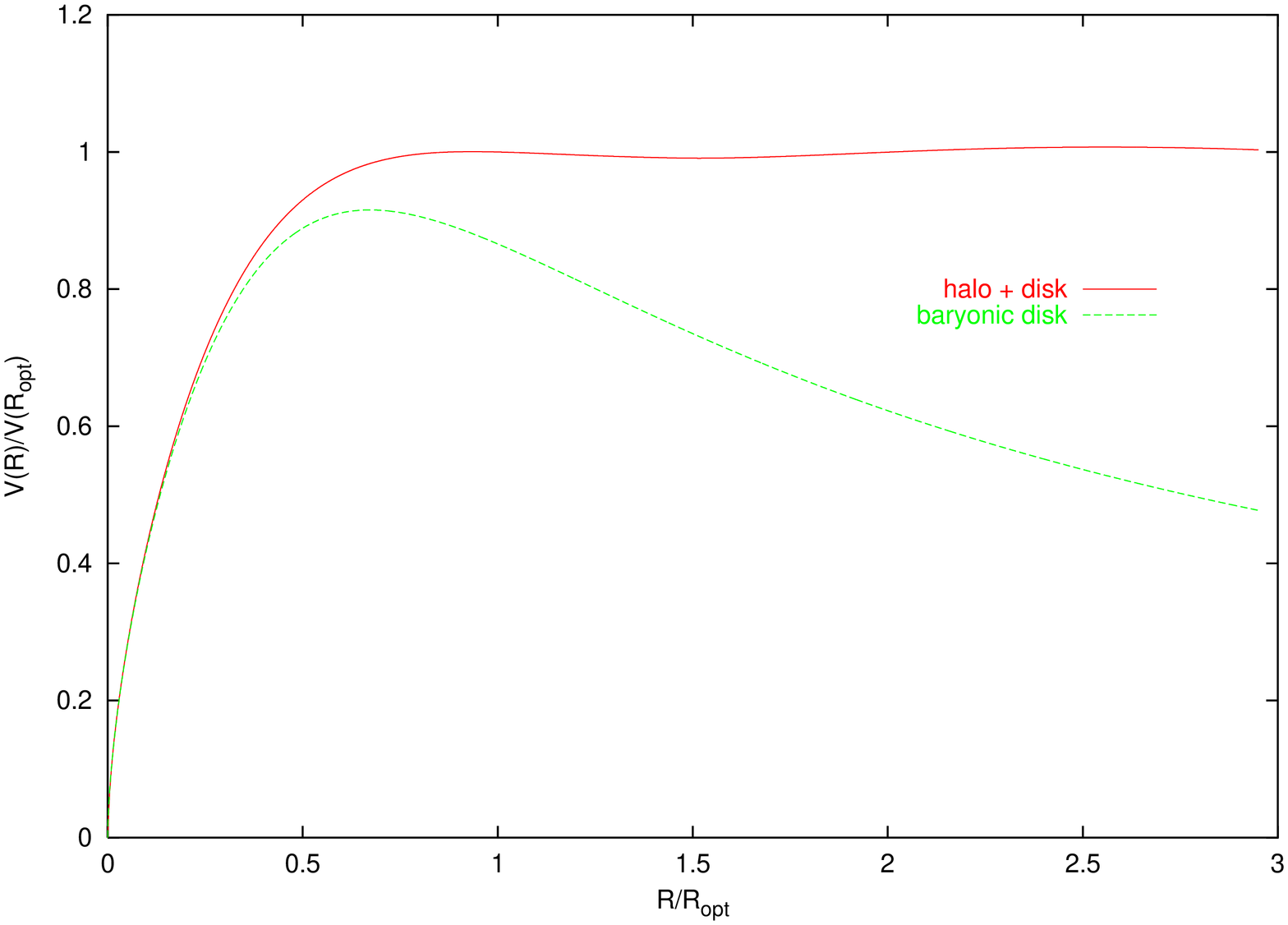}\includegraphics[width=5.8cm,angle=270]{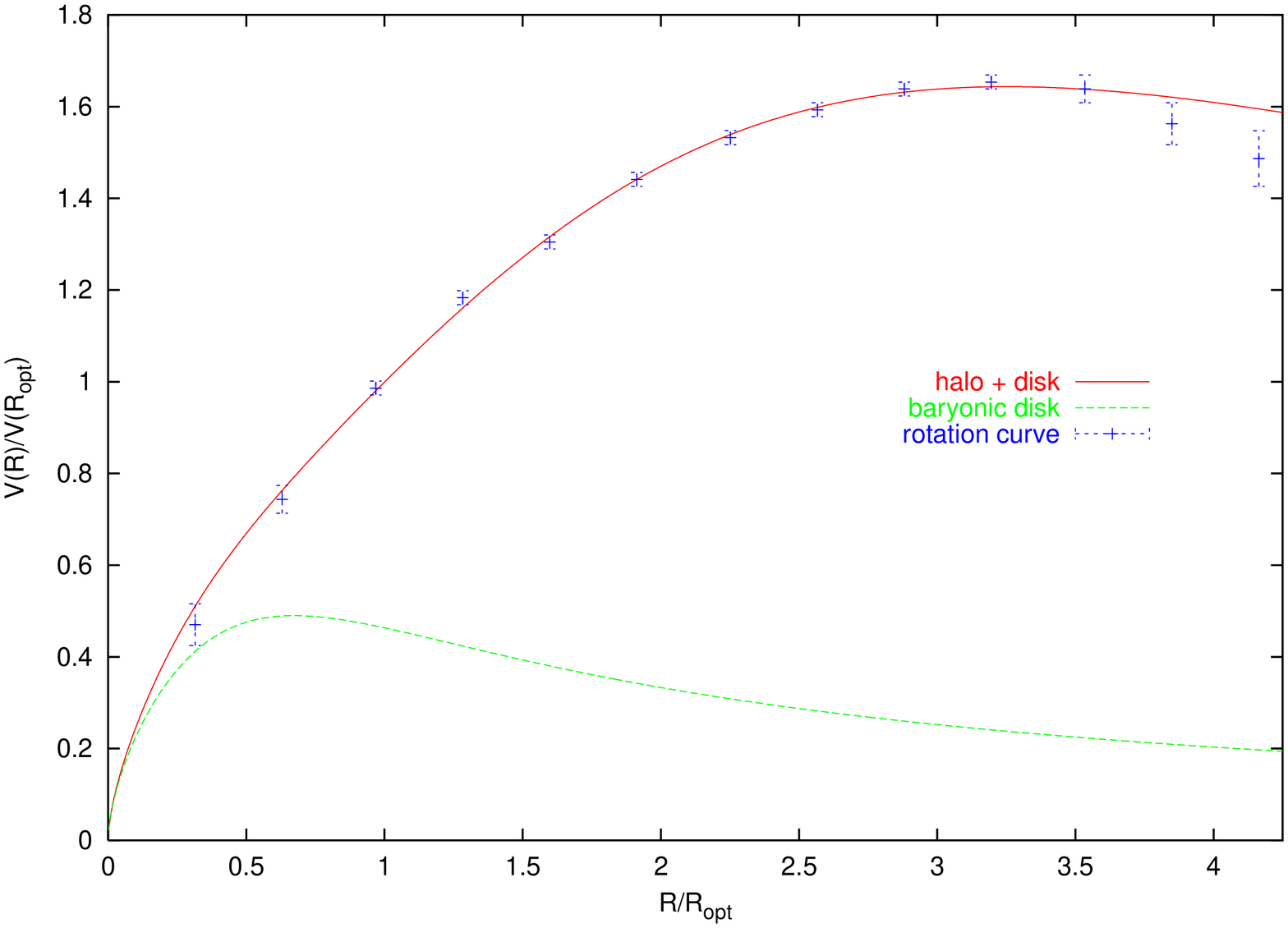}
\caption{In the left, the flat rotation curve of a fictive galaxy is shown (solid line), induced by the presence of a complex scalar field with a quadratic potential. The dashed line corresponds to the contribution from baryonic matter only. In the right, the rotation curve of the dwarf spiral DDO 154 reproduced by the same massive complex scalar field is given.}
\label{gal}
\end{figure}

The same problem appears when trying to build a dark fluid model. The potential choice is however more constrained for the dark fluid model, as the scalar field has to behave matter-like at local scale to account for observations in galaxies and clusters, and dark energy-like at cosmological scales. In \cite{Arbey:2001qi} we have already shown that it is possible to describe dark matter in galaxies and at cosmological scales using a complex scalar field, provided the scalar field potential has a mass term. We have shown in particular that with a model based on a complex scalar field $\phi$ with an internal rotation $\phi = \sigma e^{i\omega t}$ and based on the Lagrangian density:
\begin{equation}
{\cal L} \; = \;g^{\mu \nu} \, \partial_{\mu} \phi^{\dagger} \, \partial_{\nu} \phi\; - \; V \left( |\phi| \right) \;\; ,
\end{equation}
associated to a quadratic potential $V(\phi) =m^2 \phi^\dagger \phi$, flat rotation curves of spiral galaxies as well as rotation curves of dwarf spirals can be reproduced, as illustrated in Fig.~\ref{gal}. To retain the dark matter behaviour, we will consider a dark fluid potential with a $m^2 \phi^\dagger \phi$ term.

Scalar fields have shown a great versatility, as they can lead to both a dark energy behaviour and a matter behaviour. Therefore, it is quite natural to model a dark fluid using also a scalar field. However, an important question remains: how can the behaviour of the dark fluid be similar to dark matter at local scales and similar to dark energy at cosmological scales? Recalling that the density of dark fluid at cosmological scales today is of the order of the critical density, i.e. $\rho^0_c \approx 9\times10^{-29}\mbox{ g.cm}^{-3}$, and that the average matter density in galaxies is of the order $\rho^{\mbox{gal}} \approx 5 \times 10^{-24}\mbox{ g.cm}^{-3}$, a way out would be to consider that the dark fluid is inhomogeneous.

In \cite{Arbey:2006it}, the dark fluid model using the potential
\begin{equation}
V(\phi)=m^2 |\phi|^2 + A e^{-B |\phi|^2} \label{potential}
\end{equation}
was investigated. This potential is composed of a quadratic term giving a mass to the field, and mostly responsible for the dark matter behaviour, and a decreasing exponential part, which can originate from high energy theories and recalls quintessence potentials, responsible for the dark energy behaviour. In the regions of spacetime where the scalar field density is large enough, for example around galaxies or in the Early Universe, the quadratic part of the potential would dominate, and where the density is small, the second part of the potential dominates, leading to a repulsive vacuum energy-like behaviour. Thus, such a potential leads to a Universe highly inhomogeneous today.

\begin{figure}[t]
\centering
\includegraphics[width=5.8cm,angle=270]{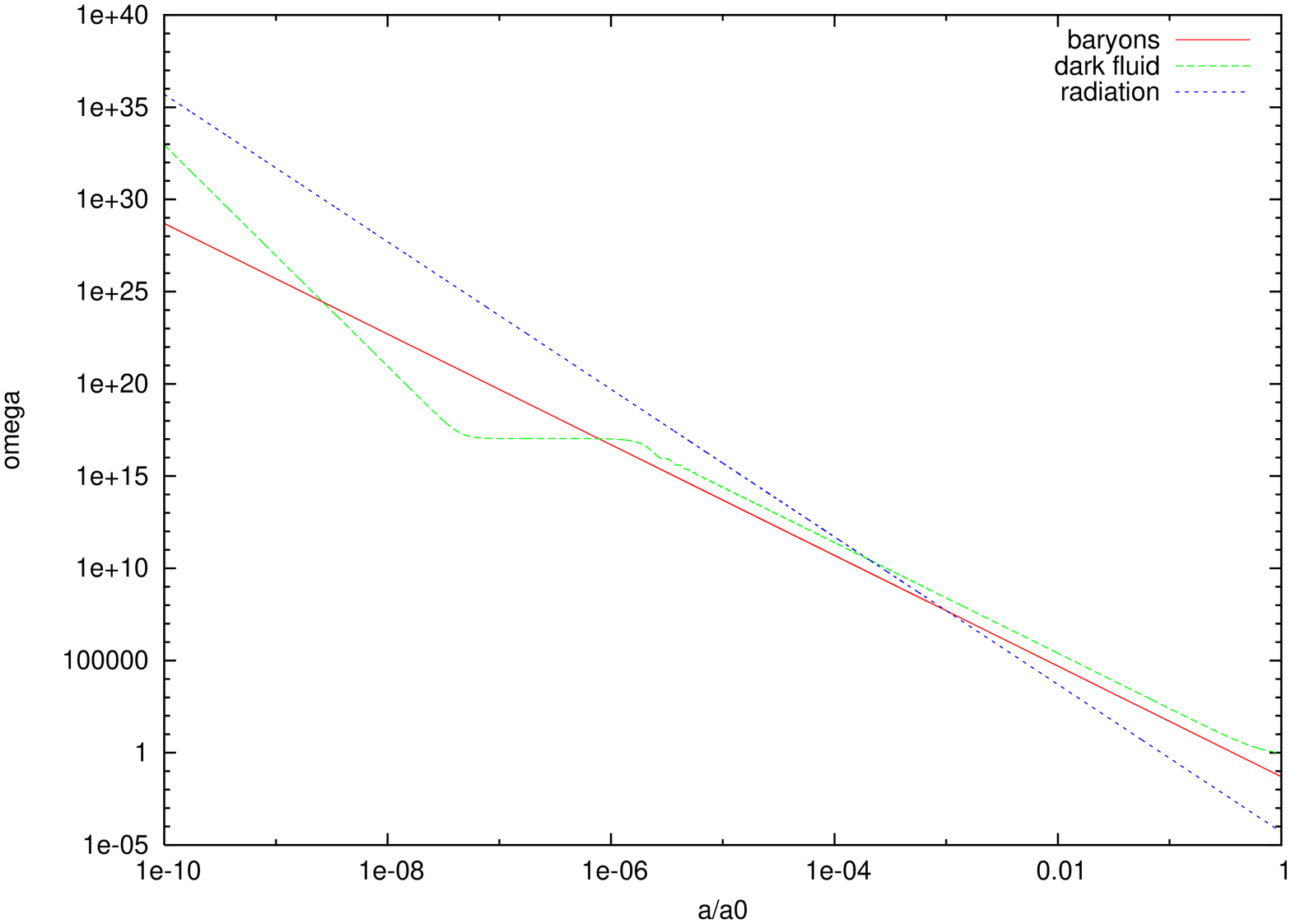}\includegraphics[width=5.8cm,angle=270]{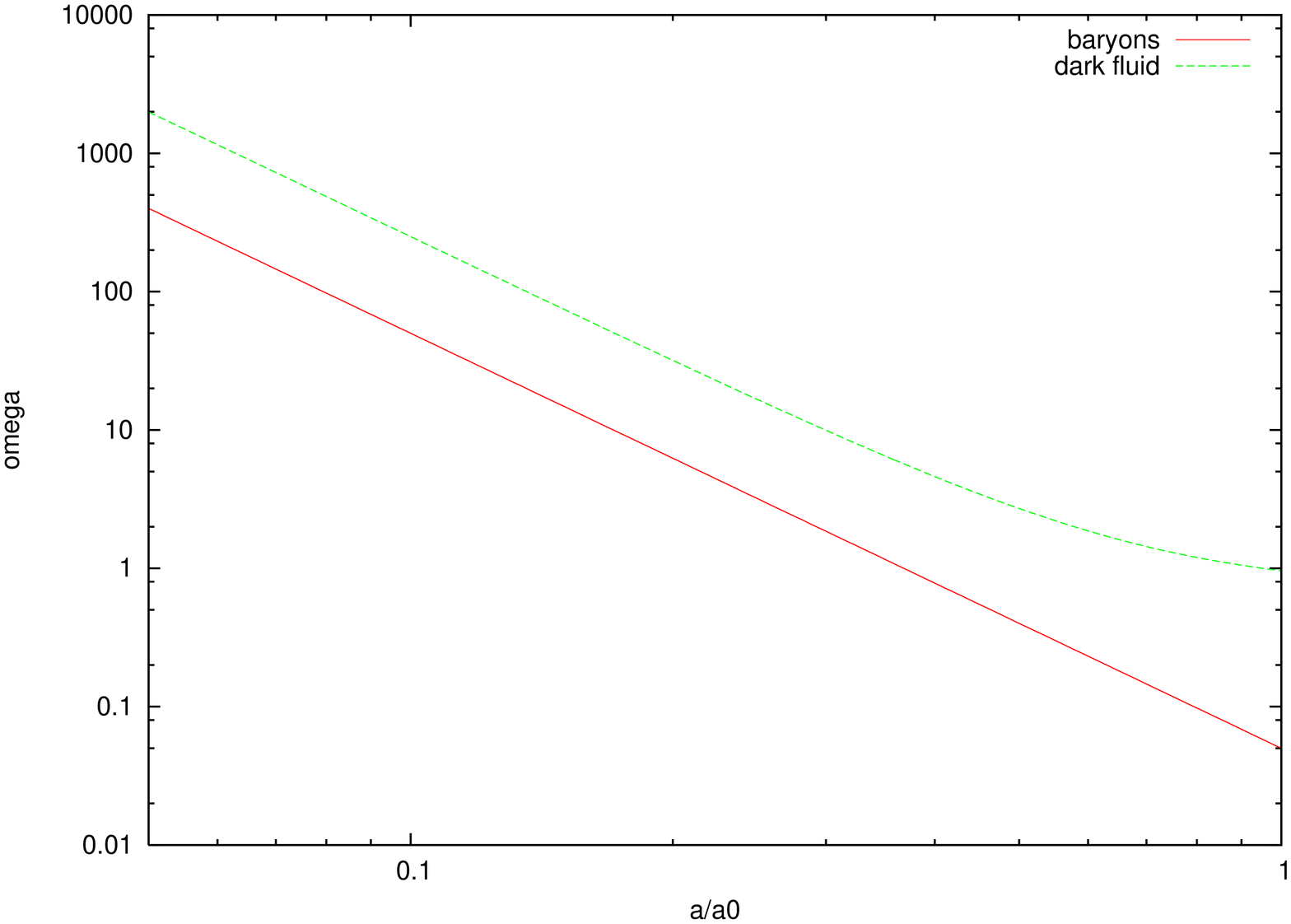}
\caption{Cosmological evolution of the density of the dark fluid scalar field in comparison to the densities of baryonic matter and radiation.}
\label{cosmo}
\end{figure}

It was also shown that such a potential can lead to the cosmological behaviour illustrated in Fig.~\ref{cosmo}, which is consistent with the observations, provided the different parameters are chosen adequately.
To fix the parameters, we consider three scales: the mass $m$ is fixed by galaxy scales, confronting the model predictions to galaxy rotation curves; the $B$ parameter is fixed in order for the field to behave mostly as dark matter at typical cluster scales; and $A$ is chosen in order for the field to be consistent with the cosmological observations revealing a dark energy behaviour. We obtain $m \sim 10^{-23}$ eV, $B \sim 10^{-22} \mbox{ eV}^{-2}$ and $A \sim \rho_0^{dark\, energy}$. With this choice of parameters, the scalar field dark fluid model is able to reproduce galaxy rotation curves, to condense at cluster scales, and to have today a negative pressure at cosmological scales.\\

\section{Quantum corrections to the dark fluid}
\begin{figure}[t]
$\begin{array}{cc}
 \includegraphics[height=4cm]{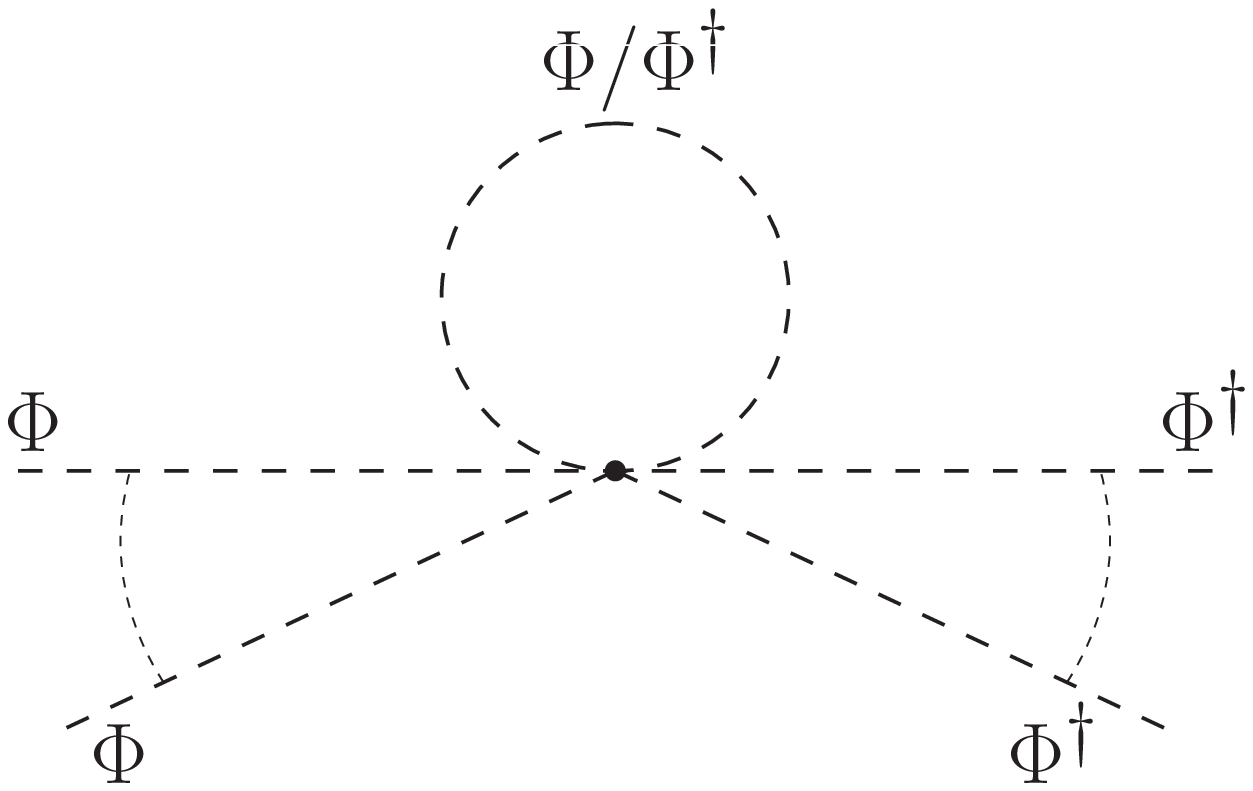}&\\[-3.cm]
&\includegraphics[height=2cm]{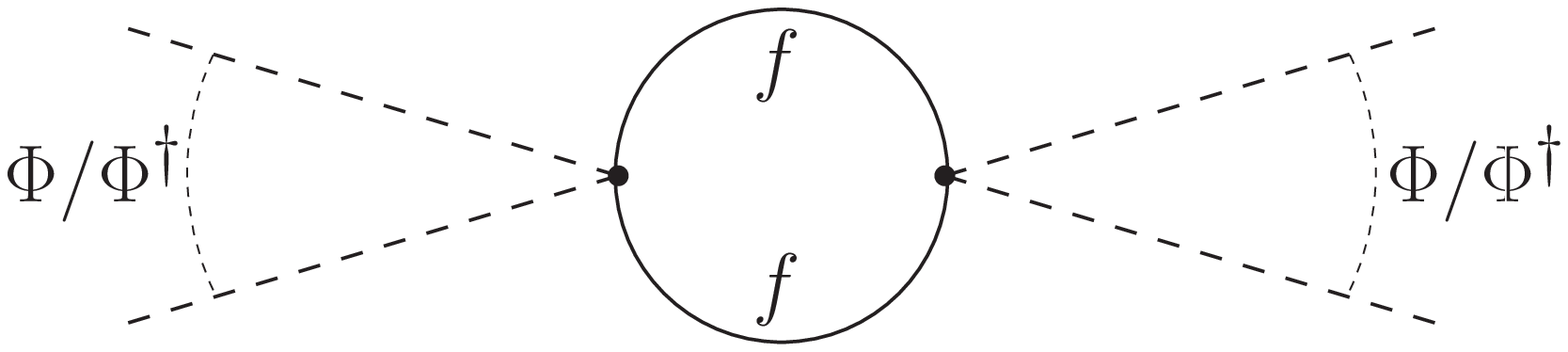}\\[0.8cm]
\end{array}$
\caption{One-loop corrections to the potential. The dashed lines represent scalar fields, and the plain lines fermions. The diagram on the left illustrates scalar corrections, whereas the right diagrams shows fermionic corrections. The multiple external lines correspond to arbitrary powers of $\Phi$ and $\Phi^\dagger$ in the potential.}
\label{fig}
\end{figure}

The question of the choice of the potential remains crucial. We would like it to be motivated by physics, and to originate if possible from high energy theories. To investigate the behaviour of the scalar field from the quantum physics point of view, we used in \cite{Arbey:2007vu} an effective quantum field theory approach to determine how quantum fluctuations would affect the potential of Eq. (\ref{potential}). The cosmological action is written as:
\begin{equation}
\mathcal{S} = \int \sqrt{|g|} \, d^4x \, \mathcal{L} \;\;,
\end{equation}
where $\mathcal{L}$ is the associated Lagrangian density. We consider that the Lagrangian contains a curvature term leading to Einstein equations, a term involving scalar fields minimally coupled to gravity, and a term containing fermions potentially coupled to the scalar fields \cite{Doran:2002bc}. Therefore
\begin{equation}
\mathcal{L} = \mathcal{L}_{curv}+\mathcal{L}_{scalar}+\mathcal{L}_{fermion} \;\;.
\end{equation}
The curvature Lagrangian is
\begin{equation}
\mathcal{L}_{curv}=M_P^2 R \;\;,
\end{equation}
where $M_P$ is the Planck mass and $R$ the curvature tensor. For a model containing a complex scalar field $\Phi$, the scalar field Lagrangian can be written as:
\begin{equation}
\mathcal{L}_{scalar}=g^{\mu\nu} \partial_\mu\Phi^\dagger(x) \partial_\nu\Phi(x)-V(\Phi(x)) \;\;,\label{Ls1}
\end{equation}
where $V$ is the potential of the field. For two real scalar fields $\phi_1$ and $\phi_2$, the scalar field Lagrangian reads:
\begin{equation}
\mathcal{L}_{scalar}=\frac{1}{2} g^{\mu\nu} [\partial_\mu\phi_1(x) \partial_\nu\phi_1(x) + \partial_\mu\phi_2(x) \partial_\nu\phi_2(x)] -V(\phi_1(x),\phi_2(x)) \;\;.\label{Ls2}
\end{equation}
Eqs. (\ref{Ls1}) and (\ref{Ls2}) are equivalent, since the complex scalar field can be written in function of the two real scalar fields:
\begin{equation}
\Phi = \frac{1}{\sqrt{2}}(\phi_1+i\phi_2) \;\;.
\end{equation}
In presence of a single fermionic species coupled to the scalar fields, the fermionic Lagrangian reads:
\begin{equation}
\mathcal{L}_{fermion}=\bar{\Psi}^\dagger(x) [i \gamma^\mu \nabla_\mu - \gamma^5 m_f(\Phi)] \Psi(x)\;\;,
\end{equation}
where $m_f(\Phi)$ is the fermion mass, which is dependent on $\Phi$, $\gamma$'s are Dirac matrices and $\nabla_\mu$ is the covariant derivative.\\
We first restrict ourselves to the flat Minkowski space, so that $g^{\mu\nu}\rightarrow\eta^{\mu\nu}$, $\sqrt{|g|}\rightarrow 1$, $R\rightarrow 0$ and $\nabla_\mu\rightarrow\partial_\mu$.\\
The method we use here to study the quantum corrections is the saddle point expansion described in \cite{Peskin:1995ev,Cheng:1985bj}. The effective action of the scalar field reads, to higher order:
\begin{equation}
\Gamma[\Phi_{cl}]= \int d^4x \, \mathcal{L}_1[\Phi_{cl}] + \frac{i}{2} \log \det\left[\frac{\delta^2 \mathcal{L}_1}{\delta \phi_i \, \delta\phi_j}\right] + \cdots \;\;,
\end{equation}
where $\mathcal{L}_1$ is the renormalized part of $\mathcal{L}_{scalar}$, the subscript $cl$ stands for classical, and $i,j=1,2$. The effective potential can be written in function of $\Gamma[\Phi_{cl}]$ such that
\begin{equation}
V_{\mbox{\small eff}}(\Phi_{cl})=-\frac{1}{VT}\Gamma[\Phi_{cl}] \;\;,
\end{equation}
where $VT$ is a space-time volume. We consider here only the one-loop corrections to the potential as shown in Fig~\ref{fig}. %
We define the effective masses
\begin{equation}
m^2_{ij}=m^2_{ji}=\frac{\partial^2 V}{\partial \phi_i \partial \phi_j} \;\;,
\end{equation}
and
\begin{eqnarray}
m_a^2 &=& \frac{1}{2} \left(m_{11}^2 + m_{22}^2 + \sqrt{(m_{11}^2 - m_{22}^2)^2 + 4 m_{12}^4}\right)\;\;,\\
m_b^2 &=& \frac{1}{2} \left(m_{11}^2 + m_{22}^2 - \sqrt{(m_{11}^2 - m_{22}^2)^2 + 4 m_{12}^4}\right) \;\;.
\end{eqnarray}
With a large momentum cutoff $\Lambda$, and ignoring the $\Phi$-independent contributions and the graphs of higher orders, the calculation of the effective potential due to the leading order scalar loop for two real coupled scalar fields leads to: 
\begin{equation}
\label{Vsca_tot}
V_{eff}(\Phi_{cl})=V(\Phi_{cl})+\frac{\Lambda^2}{32 \pi^2}(m_a^2 + m_b^2) +\frac{m_a^4 }{32\pi^2} \left[\ln\left(\frac{m_a^2}{\Lambda}\right)-\frac{3}{2}\right] + \frac{m_b^4 }{32\pi^2} \left[\ln\left(\frac{m_b^2}{\Lambda}\right)-\frac{3}{2}\right]\;\;.
\end{equation}
If the terms proportional to $m_a^4$ and $m_b^4$ are important in the context of usual field theory, in the case of cosmological scalar fields the potential is of the order of $10^{-123}$ $M_P^4$, and therefore we can safely disregard these terms in comparison to the terms proportional to $m_a^2$ and $m_b^2$. For this reason, even if many of the cosmological potentials are non-renormalizable in the strict sense of field theory, we can consider that the cosmological potentials are renormalizable as the higher order terms are so small. Therefore, the effective potential reduces to:
\begin{equation}
V_{1-loop}(\Phi_{cl})=V(\Phi_{cl})+\frac{\Lambda^2}{32 \pi^2} \left(\frac{\partial^2V}{\partial\phi_1^2}(\phi_{1\,cl},\phi_{2\,cl})+\frac{\partial^2V}{\partial\phi_2^2}(\phi_{1\,cl},\phi_{2\,cl})\right) \;\;.
\end{equation}
We can also calculate the correction due to fermions:
\begin{equation}
\label{Vfer_tot}
\delta V_{fermion}(\Phi_{cl})=-\frac{\Lambda_{ferm}^2}{8 \pi^2}[m_f(\Phi_{cl})]^2 - \frac{[m_f(\Phi_{cl})]^4 }{32\pi^2} \left[\ln\left(\frac{[m_f(\Phi_{cl})]^2}{\Lambda_{ferm}}\right)-\frac{3}{2}\right] \;\;.
\end{equation}
Again, because the fermion mass is very small in comparison to the Planck mass, the terms proportional to $[m_f(\Phi_{cl})]^4$ can be neglected. We obtain finally the one-loop effective potential for two real coupled scalar fields:
\begin{equation}
V_{1-loop}(\phi_{1\,cl},\phi_{2\,cl})=V(\phi_{1\,cl},\phi_{2\,cl})+\frac{\Lambda^2}{32 \pi^2} \left(\frac{\partial^2V}{\partial\phi_1^2}(\phi_{1\,cl},\phi_{2\,cl})+\frac{\partial^2V}{\partial\phi_2^2}(\phi_{1\,cl},\phi_{2\,cl})\right)-\frac{\Lambda_{f}^2}{8 \pi^2}[m_f(\phi_{1\,cl},\phi_{2\,cl})]^2 \;\;,
\end{equation}
or, for a single complex scalar field:
\begin{equation}
V_{1-loop}(\Phi_{cl})=V(\Phi_{cl})+\frac{\Lambda^2}{64 \pi^2} \frac{\partial^2V}{\partial\Phi^\dagger \partial\Phi}(\Phi_{cl})-\frac{\Lambda_{f}^2}{8 \pi^2}[m_f(\Phi_{cl})]^2 \;\;.
\end{equation}
These results are in agreement with the single real scalar field results of \cite{Doran:2002bc}.\\
We disregarded here the higher order terms, which is a reasonable approximation, as the potential and its derivatives are today of the order of the critical density, {\it i.e.} of the order of $10^{-123} M_P^4$. We also note that the ignored $\Phi$-independent contributions would lead to a cosmological constant of the order $\Lambda^4=\mathcal{O}(M_P^4)$, which is by far much larger than the critical density. This is a well-known problem appearing in the majority of field theories and which still has to be solved.\\
In the following, we consider that $\Lambda = M_P$ unless stipulated otherwise, and we use units in which $c = \hbar = M_P = 1$. We consider the potential \cite{Arbey:2006it}:
\begin{equation}
V(\Phi)=m^2 \Phi^\dagger \Phi + \alpha \exp(-\beta \Phi^\dagger \Phi) \;\;.\label{DF1}
\end{equation}
This potential can lead to both dark matter and dark energy behaviours. The one-loop effective potential reads:
\begin{equation}
V_{1-loop}= \frac{\Lambda^2}{64\pi^2}m^2 + m^2 \Phi_{cl}^\dagger\Phi_{cl} + \frac{\Lambda^2}{64\pi^2}\alpha \beta^2 \ \Phi_{cl}^\dagger\Phi_{cl} \exp(-\beta \Phi_{cl}^\dagger\Phi_{cl}) + \alpha \left(1 - \beta \frac{\Lambda^2}{64\pi^2} \right) \exp(-\beta \Phi_{cl}^\dagger\Phi_{cl}) \;\;.
\end{equation}
Again, the constant term can be safely disregarded. The term $-\beta \Lambda^2/(64\pi^2) \exp(-\beta \Phi_{cl}^\dagger\Phi_{cl})$ can be reabsorbed in the redefinition $\alpha \rightarrow \alpha/\left(1 - \beta\Lambda^2/(64\pi^2) \right)$. However, a problem arises from the extra term $\Lambda^2/(64\pi^2) \alpha \beta^2 \Phi_{cl}^\dagger\Phi_{cl} \exp(-\beta \Phi_{cl}^\dagger\Phi_{cl})$ which modifies the potential. Considering the values of the parameters derived in \cite{Arbey:2006it}, this term cannot be neglected today and would lead to a strong modification of the potential, unless $\Lambda \ll 10^{-4}$. We can however note that at earlier epochs, as $|\Phi|$ was larger, this term was suppressed by the exponential.

A way to solve this problem would be to modify the potential such as:
\begin{equation}
V(\Phi)=m^2 \Phi^\dagger \Phi + (A + B \Phi^\dagger \Phi) \exp(-\beta \Phi^\dagger \Phi) \;\;.
\end{equation}
In this case, the effective potential reads:
\begin{eqnarray}
V_{1-loop} &=& \frac{\Lambda^2}{64\pi^2}m^2 + m^2 \Phi^\dagger \Phi  + \frac{\Lambda^2}{64\pi^2} B \beta^2 (\Phi_{cl}^\dagger\Phi_{cl})^2 \exp(-\beta \Phi_{cl}^\dagger\Phi_{cl})  \\ 
&& + \left[ \left(A + (B -  A \beta) \frac{\Lambda^2}{64\pi^2} \right) + \left(B + \frac{\Lambda^2}{64\pi^2} (A \beta^2 - 3 B \beta)\right) \Phi_{cl}^\dagger\Phi_{cl} \right] \exp(-\beta \Phi_{cl}^\dagger\Phi_{cl})\nonumber\;\;.
\end{eqnarray}
The constant term is irrelevant. With two redefinitions: $A \rightarrow A - (B -  A \beta) \Lambda^2/(64\pi^2)$ and $B \rightarrow B - (A \beta^2 - 3 B \beta) \Lambda^2/(64\pi^2) $, the correction terms of the second line vanish, but the last correction term of the first line still remains and modifies the shape of the potential. However, for $B \sim 10^{-120}$, this term is negligible in comparison to the other terms, and the shape of the potential is then conserved. If $A=\alpha$ and $B$ is small in comparison to $m^2$, the analysis of \cite{Arbey:2006it} remains valid. In the case of a larger $B$, we could modify again the potential, such as:
\begin{equation}
V(\Phi)=m^2 \Phi^\dagger \Phi + \sum_n \bigl(\alpha_n (\Phi^\dagger \Phi)^n \bigr)\exp(-\beta \Phi^\dagger \Phi) \;\;, \label{DF2}
\end{equation}
where $n$ goes from 0 to $n_{max}$. In this case, the corrections can be reabsorbed into the constants $\alpha_n$, and the last correction of order $(\Phi_{cl}^\dagger \Phi_{cl})^{n_{max}+1}$ will be negligible today in comparison to the other terms. Also, such a potential would lead to a behaviour similar to that described in \cite{Arbey:2006it}, provided the redefined terms $\alpha_n (\Phi_{cl}^\dagger \Phi_{cl})^n$ (for $n \ne 0$) are today small in comparison to $\alpha_0$.

\noindent We will now consider the corrections due to the coupling of the scalar fields to fermions. The fermion term in the effective potential reads:
\begin{equation}
V_{1-loop}^{f}(\Phi_{cl})=-\frac{\Lambda_{f}^2}{8 \pi^2}[m_f(\Phi_{cl})]^2 \;\;.
\end{equation}
If the fermion mass is $\Phi$-independent and of the order of 100 GeV, and considering that the fermionic cutoff is taken at GUT scale: $\Lambda_{f}=10^{-3}$, the fermionic term is of the order of $10^{-42}$, which is extremely large in comparison to the present value of the potential $V(\Phi_{cl}) \sim 10^{-123}$. Thus, as noticed in \cite{Doran:2002bc}, the only way that the fermionic corrections do not strongly modify the potential is that $[m_f(\Phi_{cl})]^2$ takes a form which is already contained in the classical potential, and can therefore be reabsorbed in the constants of the potential.

Let us focus for example on the dark fluid potential of Eq. (\ref{DF2}). The fermion mass can then take a form such as:
\begin{equation}
m_f(\Phi_{cl})=m_f^0 (\Phi_{cl}^\dagger \Phi_{cl})^{n/2} \exp\left(-\frac{1}{2} \beta \Phi_{cl}^\dagger\Phi_{cl}\right) \;\;,
\end{equation}
or the square root of a sum of similar terms. However, a simpler acceptable form would be:
\begin{equation}
[m_f(\Phi_{cl})]^2=(m_f^0)^2 \Phi_{cl}^\dagger \Phi_{cl} \;\;.
\end{equation}
In this case, $(m_f^0)^2$ can be reabsorbed in the redefinition $m^2 \rightarrow m^2 + (m_f^0)^2$.

We now consider the case where the coupling is not of the shape of the potential, and where the one-loop corrections can not be reabsorbed by higher order corrections. We can write the fermion mass as
\begin{equation}
m_f(\Phi_{cl})= m_f^0 + \delta m_f(\Phi_{cl}) \;\;.
\end{equation}
Assuming that $\delta m_f(\Phi_{cl}) \ll m_f^0$, we have:
\begin{equation}
V_{1-loop}^{f}(\Phi_{cl})=-\frac{\Lambda_{f}^2}{8 \pi^2}[(m_f^0)^2+ 2 m_f^0 \delta m_f(\Phi_{cl}) + \delta m_f(\Phi_{cl})^2] \;\;.
\end{equation}
The constant term is irrelevant, as the effective potential is defined up to a constant. Thus, the global potential has a similar structure if
\begin{equation}
\frac{\Lambda_{f}^2}{4 \pi^2} m_f^0 \delta m_f(\Phi_{cl}) \ll V(\Phi_{cl}) \;\;,
\end{equation}
For a present value of the potential such as $V(\Phi_{cl}) \sim 10^{-123}$, this leads to a limit on the present value of $\delta m_f$:
\begin{equation}
\delta m_f(\Phi_{cl}) \ll 10^{-79} \mbox{ GeV} \;\;.
\end{equation}
This limit is so stringent that it nearly forbids a dependence of the mass on $\Phi_{cl}$ if the shape of the coupling is different from the shape of the potential. This result is similar for the quintessence potentials, and is in agreement with \cite{Doran:2002bc}. We however have to notice that as we restricted ourselves to one-loop corrections, this limit has to be interpreted with reserve, as the higher order corrections could also modify the structure of the effective potential.

We have shown that the original potential (\ref{DF1}) would be modified by quantum fluctuations, especially if we expect it to be somehow non-minimally coupled to matter fields. This result can be understood in two different manners: either such a potential is not resistant to quantum fluctuation and is not viable from the quantum theory point of view, or it is an effective potential which already includes the quantum corrections.

\section{Conclusion}
\noindent The standard model of cosmology assumes the existence of two unknown dark components, dark matter and dark energy. We have seen that it is possible to replace both components with a unique component, the dark fluid, and be consistent with the cosmological data. The properties of the dark fluid are severely constrained because of the dual dark energy/dark matter conditions, but it is nevertheless possible to model the dark fluid with a simple and usual complex scalar field. The main question of the model is the choice of the scalar field potential. An adequate choice can lead to a fluid explaining at the same time the dark matter observations at local scales, and the dark energy behaviour at cosmological scales. The possibility that a dark fluid can lead to a correct structure formation has still to be tested, and methods to simulate scalar field condensation are being developed \cite{Bernal:2006ci,Bernal:2009zy}. However, the relations between such a model and quantum theories are not obvious, and still need to be worked out. In this context, a new interesting potential is currently under investigation:
\begin{equation}
V(\Phi) = \alpha \, \mbox{cotanh}\left( \frac{\beta}{\Phi^\dagger \Phi} \right)\;\;.
\end{equation}
It has the advantage of finding roots in brane theories, and yet no big difference in behaviour is expected from the one determined by the potential in Eq. (\ref{potential}). Another way to find an adequate potential would be to study the possibility of an even larger unification, {\it i.e.} dark matter, dark energy and inflation, as proposed in \cite{Liddle:2006qz,PerezLorenzana:2007qv,Liddle:2008bm,Bose:2008ew}.

To conclude, dark fluid models appear as viable alternatives to the standard cosmological model, and need to be further investigated.

%%%%%%%%%%%%%%%%%%%%%%%%%%%%%%%%%%%%%%%%%%%%%%%%
%% BACKMATTER
%%%%%%%%%%%%%%%%%%%%%%%%%%%%%%%%%%%%%%%%%%%%%%%%

\bibliographystyle{aipproc}   % if natbib is available

%\bibliography{darkfluid_arbey}

\begin{thebibliography}{22}
\expandafter\ifx\csname natexlab\endcsname\relax\def\natexlab#1{#1}\fi
\providecommand{\enquote}[1]{``#1''}
\expandafter\ifx\csname url\endcsname\relax
  \def\url#1{\texttt{#1}}\fi
\expandafter\ifx\csname urlprefix\endcsname\relax\def\urlprefix{URL }\fi
\providecommand{\eprint}[2][]{\url{#2}}

\bibitem[Arbey(2005)]{Arbey:2005fn}
A.~Arbey (2005), \eprint{astro-ph/0506732}.

\bibitem[Komatsu et~al.(2009)]{Komatsu:2008hk}
E.~Komatsu, et~al., \emph{Astrophys. J. Suppl.} \textbf{180}, 330--376 (2009),
  \eprint{0803.0547}.

\bibitem[Arbey(2008)]{Arbey:2008gw}
A.~Arbey, \emph{Open Astron. J.} \textbf{1}, 27--38 (2008), \eprint{0812.3122}.

\bibitem[Arbey and Mahmoudi(2008)]{Arbey:2008kv}
A.~Arbey, and F.~Mahmoudi, \emph{Phys. Lett.} \textbf{B669}, 46--51 (2008), \eprint{0803.0741}; A.~Arbey, and F.~Mahmoudi (2009{\natexlab{a}}), \eprint{0906.0368}; A.~Arbey, and F.~Mahmoudi (2009{\natexlab{b}}), \eprint{0909.0266}.

\bibitem[Kawano(1992)]{Kawano:1992ua}
L.~Kawano  (1992), FERMILAB-PUB-92-004-A.

\bibitem[Jedamzik(2006)]{Jedamzik:2006xz}
K.~Jedamzik, \emph{Phys. Rev.} \textbf{D74}, 103509 (2006),
  \eprint{hep-ph/0604251}.

\bibitem[Kamenshchik et~al.(2001)]{Kamenshchik:2001cp}
A.~Y. Kamenshchik, U.~Moschella, and V.~Pasquier, \emph{Phys. Lett.}
  \textbf{B511}, 265--268 (2001), \eprint{gr-qc/0103004}.

\bibitem[Bean and Dore(2003)]{Bean:2003ae}
R.~Bean, and O.~Dore, \emph{Phys. Rev.} \textbf{D68}, 023515 (2003),
  \eprint{astro-ph/0301308}.

\bibitem[Bilic et~al.(2002)]{Bilic:2001cg}
N.~Bilic, G.~B. Tupper, and R.~D. Viollier, \emph{Phys. Lett.} \textbf{B535},
  17--21 (2002), \eprint{astro-ph/0111325}.

\bibitem[Arbey(2006)]{Arbey:2006it}
A.~Arbey, \emph{Phys. Rev.} \textbf{D74}, 043516 (2006),
  \eprint{astro-ph/0601274}.

\bibitem[Peebles and Ratra(1988)]{Peebles:1987ek}
P.~J.~E. Peebles, and B.~Ratra, \emph{Astrophys. J.} \textbf{325}, L17 (1988).

\bibitem[Arbey et~al.(2001)]{Arbey:2001qi}
A.~Arbey, J.~Lesgourgues, and P.~Salati, \emph{Phys. Rev.} \textbf{D64}, 123528 (2001), \eprint{astro-ph/0105564}; A.~Arbey, J.~Lesgourgues, and P.~Salati, \emph{Phys. Rev.} \textbf{D65}, 083514 (2002), \eprint{astro-ph/0112324}; A.~Arbey, J.~Lesgourgues, and P.~Salati, \emph{Phys. Rev.} \textbf{D68}, 023511 (2003), \eprint{astro-ph/0301533}.

\bibitem[Arbey and Mahmoudi(2007)]{Arbey:2007vu}
A.~Arbey, and F.~Mahmoudi, \emph{Phys. Rev.} \textbf{D75}, 063513 (2007),
  \eprint{hep-th/0703053}.

\bibitem[Doran and Jaeckel(2002)]{Doran:2002bc}
M.~Doran, and J.~Jaeckel, \emph{Phys. Rev.} \textbf{D66}, 043519 (2002),
  \eprint{astro-ph/0203018}.

\bibitem[Peskin and Schroeder(1995)]{Peskin:1995ev}
M.~E. Peskin, and D.~V. Schroeder  (1995), Reading, USA: Addison-Wesley (1995)
  842 p.

\bibitem[Cheng and Li(1984)]{Cheng:1985bj}
T.~P. Cheng, and L.~F. Li  (1984), Oxford, UK: Clarendon (1984) 536 P. (Oxford
  Science Publications).

\bibitem[Bernal and Siddhartha~Guzman(2006)]{Bernal:2006ci}
A.~Bernal, and F.~Siddhartha~Guzman, \emph{Phys. Rev.} \textbf{D74}, 103002
  (2006), \eprint{astro-ph/0610682}.

\bibitem[Bernal et~al.(2009)]{Bernal:2009zy}
A.~Bernal, J.~Barranco, D.~Alic, and C.~Palenzuela  (2009), \eprint{0908.2435}.

\bibitem[Liddle and Urena-Lopez(2006)]{Liddle:2006qz}
A.~R. Liddle, and L.~A. Urena-Lopez, \emph{Phys. Rev. Lett.} \textbf{97},
  161301 (2006), \eprint{astro-ph/0605205}.

\bibitem{PerezLorenzana:2007qv}
  A.~Perez-Lorenzana, M.~Montesinos, and T.~Matos,
  \emph{Phys. Rev.}  \textbf{D77}, 063507 (2008),
  \eprint{0707.1678}.

\bibitem[Liddle et~al.(2008)]{Liddle:2008bm}
A.~R. Liddle, C.~Pahud, and L.~A. Urena-Lopez, \emph{Phys. Rev.} \textbf{D77},
  121301 (2008), \eprint{0804.0869}.

\bibitem{Bose:2008ew}
  N.~Bose, and A.~S.~Majumdar, \emph{Phys. Rev.} \textbf{D79}, 103517 (2009), \eprint{0812.4131}.

\end{thebibliography}

\hyphenation{Post-Script Sprin-ger}

\end{document}